\documentclass[conference]{IEEEtran} 

\usepackage[dvips]{graphicx}
\usepackage{subcaption}
\usepackage{epsf}
\usepackage{epsfig}
\usepackage[noadjust]{cite}
\usepackage{amsmath}
\usepackage{url}
\usepackage[export]{adjustbox}
\usepackage{soul}
\usepackage{algorithm}
\usepackage[noend]{algpseudocode}
\usepackage{color}
\usepackage{tree-dvips}
\usepackage{varwidth}
\usepackage{lineno}
\usepackage{lipsum}
\usepackage{multicol}
\usepackage{fancyvrb}
\usepackage{graphicx,dblfloatfix}
\usepackage{tikz}
\usepackage{comment}
\usepackage{xcolor}

\def\BibTeX{{\rm B\kern-.05em{\sc i\kern-.025em b}\kern-.08em
    T\kern-.1667em\lower.7ex\hbox{E}\kern-.125emX}}

\begin{document}

\onecolumn 

{\LARGE IEEE Copyright Notice} \\

\copyright 2019 IEEE. Personal use of this material is permitted. Permission from IEEE must be obtained for all other uses, in any current or future media, including reprinting/republishing this material for advertising or promotional purposes, creating
new collective works, for resale or redistribution to servers or lists, or reuse of any copyrighted component of this work in other works. \\

{\large Accepted to be Published in: Proceedings of the 2019 IEEE International Midwest Symposium on Circuits and Systems (MWSCAS), Aug. 4-7, 2019, Dallas, TX, USA.}

\twocolumn

\IEEEoverridecommandlockouts

\title{Variable Record Table: A Run-time Solution for Mitigating Buffer Overflow Attack} 

\author{\IEEEauthorblockN{Love Kumar Sah, Sheikh Ariful Islam, and Srinivas Katkoori}
\IEEEauthorblockA{Department of Computer Science and Engineering\\ University of South Florida \\ Tampa, FL 33620\\ Email: \{lsah, sheikhariful, katkoori\}@mail.usf.edu}\vspace*{-0.8cm}}
 \maketitle

\begin{abstract}

We present a novel approach to mitigate buffer overflow attack using Variable Record Table (VRT). Dedicated memory space is used to automatically record base and bound information of variables extracted during runtime. We instrument frame pointer and function(s) related registers to decode variable memory space in stack and heap. We have modified Simplescalar/PISA simulator to extract variables space of six (6) benchmark suites from MiBench. We have tested 290 small C programs (MIT corpus suite) having 22 different buffer overflow vulnerabilities in stack and heap. Experimental results show that our approach can detect buffer overflow attack with zero instruction overhead with the memory space requirement up to 13Kb to maintain VRT for a program with 324 variables. 
\end{abstract}



 \section{Introduction}
 
National Vulnerability Database (NVD) records show that in the past two years, the cases of buffer overflow attack was almost thrice than previous years \cite{NIST}. Several library functions in C programming lack variable boundary checking leading to potential memory corruption unknowingly.
Many operating system kernel and drivers  written in such unsafe language allow an attacker to exploit the system. An attacker normally overflows variable(s) space to modify the control data and hijack the program control. 
 
Ever since the introduction of stack smashing \cite{Smashing} in the early 90s, the stack has been vulnerable to multiple attacks. Many software and hardware approaches have been proposed to prevent such attacks. The software-based approach  performs both static and lexical  analysis of  the  code  to  find  vulnerable  function(s),  function call, and  illegal accesses of array  element(s)  in  the  source  code  or the binary.  Techniques such as FlawFinder \cite{Flawfinder},  RATS\mbox{\cite{RATS}},  and  LibSafe \cite{libsafe} performed an exhaustive search  to  match  tokens  in  program semantics  against  a database of known vulnerabilities. However, software-only approaches limit their application to debugging purposes and incur performance overhead as high as 30X \cite{30x}. On the contrary, hardware solutions are faster and transparent to the running process.  Non-Executable stack \cite{nonex}, Address Space Layout Randomization (ASLR) \cite{ASLR}, Canary word \cite{StackGuard} next to return address have been effective with extra instruction overhead. Maintaining shadow stack \cite{ASLR} incurs performance overhead as high as 10\%. Most solutions proposed so far implement either additional lines of code to check the bound or maintain expensive memory isolation strategy. 

In this paper, we propose a novel approach to extract the memory space information by instrumenting instructions during runtime. We store this information in a table called Variable Record Table (VRT). We track frame pointer operation in instructions to extract static variables space in the stack. For heap space, we instrument the argument and return registers during a dynamic memory function call by an instruction. VRT is built upon instrumenting object file only during runtime. With this variable level information, memory related security issues can be handled by the processor itself. We use VRT entries to check out-of-bound access during buffer overflow. The novelty of the proposed work is as follows:  

\begin{enumerate}
\item Extract variable runtime memory space information.
\item Detect buffer overflow cases utilizing VRT entries.
\end{enumerate}

Experimental results for MIT Static Corpus benchmark \cite{Kartkiewicz2009} (290 C programs) on SimpleScalar toolset {\cite{SimpleScalar}} successfully detect  different buffer overflow cases. We also demonstrate that our approach detects buffer overflow with no additional instruction overhead. The memory overhead to maintain VRT a program with 324 variables was only 13Kb.

The  remainder  of  the  paper  is  organized  as  follows.  In Section II, we present background, discuss the attack on process memory. In  Section  III,  we  describe  the details of the proposed approach. In Section IV, we report the experimental results. In Section V, we draw conclusions and mention future directions.


\section{Background}
\label{sec:related}
\subsection{ Buffer Overflow}

During a buffer overflow, variables access go beyond their allocated space that may overwrite data into adjacent memory space. Such undesirable and illegal scenario pose a security threat during program execution. Fig. \ref{fig:heap_c_code} shows  an exploitable program with overflow case. In the {\tt strcpy()} function, there exists no bound checking of the destination variable, {\tt p} and unrestricted source variable data, {\tt argv[1]} provided by the user could override {\tt p}'s heap space resulting in data corruption of adjacent space.

\vspace{-2ex}
\begin{figure}[H]
\begin{varwidth}{\linewidth}
\centering
\begin{verbatim}
  int main(int argc, char **argv)
   {
    char *p, *q;
    p = malloc(1024);
    if (argc >= 2)
        strcpy(p, argv[1]);
    return 0; 
   }
\end{verbatim}
\end{varwidth}
\vspace{-1ex}
\caption{Buffer Overflow Program}
\label{fig:heap_c_code}
\end{figure}

\subsection{Process Memory}

\begin{figure}
\includegraphics[width=1.01\columnwidth]{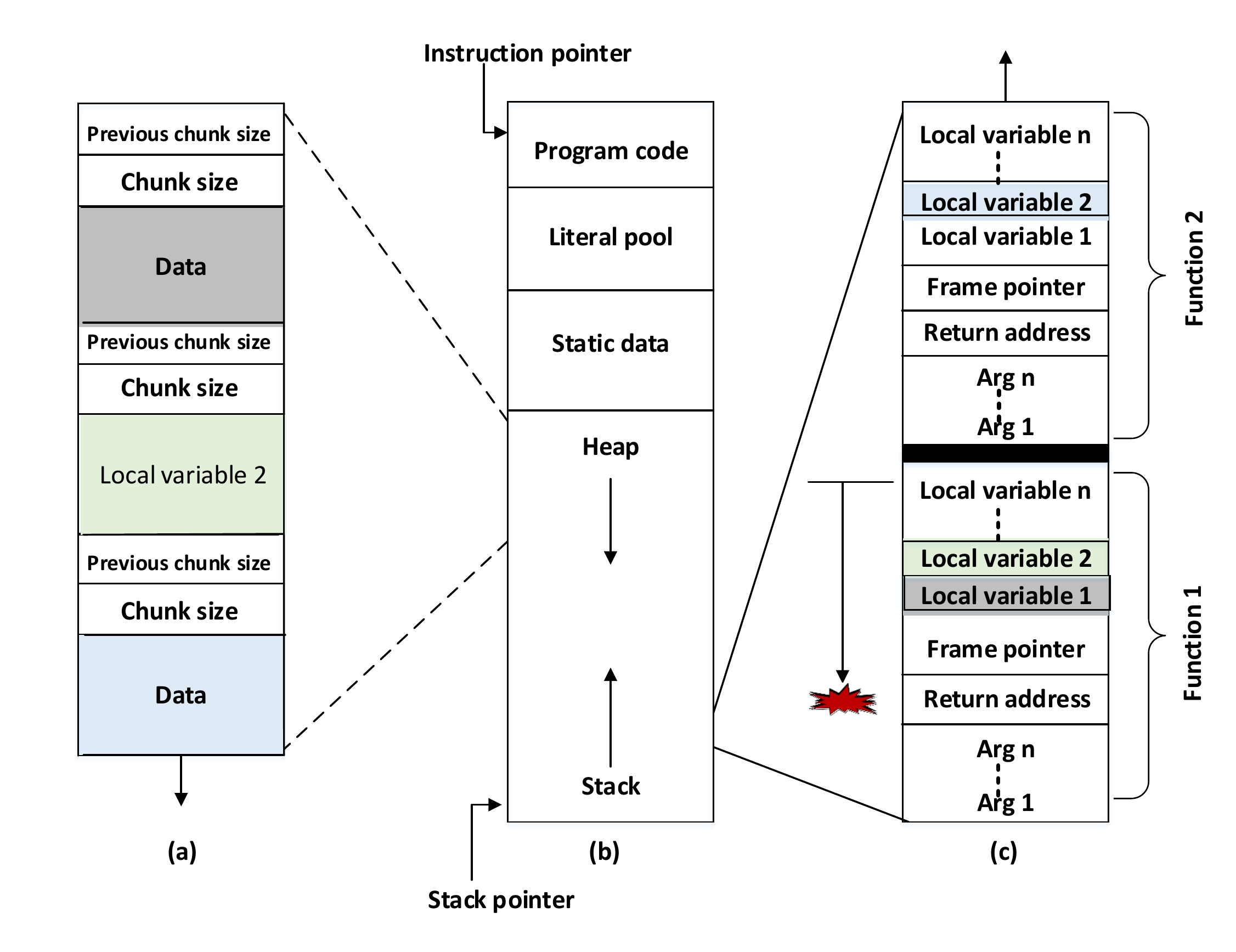}
\caption{Process Memory Organization}
\label{fig:stack_memory1}
\end{figure}

Fig. \ref{fig:stack_memory1}(b) depicts the main memory space for a running process. It consists of code and data segment for storing program instructions and data respectively. Stack memory segment in Fig. \ref{fig:stack_memory1}(c) is used to store static variables of the program. It also stores additional information of the function such as  arguments, return address, and previous frame pointer in order to maintain the  function execution. On the contrary, the heap segment in Fig. \ref{fig:stack_memory1}(a)  stores the dynamic data during the runtime. A heap  dynamic block space is linked with the static variables in stack space (shown as color coded in figure). Heap space can be dynamically allocated or freed using library functions in the program whereas the stack space is fixed for a function call.

\section{Proposed Approach and Implementation}
\label{sec:proposed}
In this section, we present  static and dynamic memory space extraction process followed by discussion on the VRT table and its format. Then, we discuss the use of VRT for different cases of buffer overflow. 

\subsection{Static Variable Space}

In the stack, static variable space for a function is laid out in sequential order by the compiler. Such variable space starts in the stack after the return address, frame pointer, saved registers, and argument(s) from the previous function. To find the base address of the local variables, we track the frame pointer's offset in  load ({\tt lw}) and/or store ({\tt sw}) addresses. The relative difference between sequential variable addresses can provide the size of the allocated space. 

\begin{figure}[h]
\centering
\begin{varwidth}{\linewidth}
\begin{verbatim}
1  int main()
2  { 
3   int my_arr[] = {13,56,71,38,93,12}; 
4   int *ptr, i;
5   ................
6  }
\end{verbatim}
\end{varwidth}
\caption{Static variable declaration example in C}
\label{fig:main_func}
\end{figure}

\begin{figure}
\centering
\begin{varwidth}{\linewidth}
\begin{verbatim}
1.   4001f0 <main>:
2.   4001f0:	addiu $29,$29,-56
3.   4001f8:	sw $31,52($29)
4. * 400200:	sw $30,48($29)
5.   400208:	addu $30,$0,$29
6.   400210:	jal 400618 <__main>
7. * 400218:	addiu $2,$30,16
  		            :         
8  * 4002b0:	sw $2,40($30)
9  * 4002b8:	sw $0,44($30)
10   400358:	addu $29,$0,$30
11   400360:	lw $31,52($29)
12   400368:	lw $30,48($29)
13   400370:	addiu $29,$29,56
14   400378:	jr $31  
\end{verbatim}
\end{varwidth}
\caption{MIPS Assembly code }
\label{fig:asm}
\vspace{-2ex}
\end{figure} 

 In Fig. \ref{fig:main_func} we have three variables (basic, array and pointer) of {\tt int} type declared whose assembly code in MIPS is shown in Fig.  \ref{fig:asm}. In this figure, once the frame pointer is initialized with the stack pointer address (line 7), it allocates the local variables' space based on their size and order of use in the program. We observe that the frame pointer with offset value 16 for the array (line 9), 40 for the pointer (line 10), and 44 for the variable {\tt i} (line 11). To find the bound, we subtract two adjacent variables' base addresses. For example, the bound value of the array will be (40-16) = 24. For the last variable (variable {\tt i}) the bound information  can be calculated by subtracting it from line 6 offset when it finishes storing the previous function data.

\subsection{Dynamic Variable Space}

Dynamic variable space is reserved or freed using the library functions. {\tt malloc()}, {\tt realloc()}, {\tt free()} and their variants  are Direct Memory Access (DMA) functions in C. We present the technique to extract the variables space from the assembly code of these functions.

 malloc(): This function takes an integer and return heap space. 
 
\vspace{-2ex}
\begin{figure}[H]
\centering
\begin{varwidth}{\linewidth}
\begin{verbatim}
int *p1 = malloc(size);
\end{verbatim}
\end{varwidth}
\label{fig:hlow_c_code}

\end{figure}

Instruction level decomposition of the above {\tt malloc} function call is shown below. We notice that function call ({\tt jal}) has the argument register (\$4) with an integer value and upon return, return register (\$2) with heap address.

\vspace{-1.5ex}
\begin{figure}[h]
\centering
\begin{varwidth}{\linewidth}
\begin{verbatim}
addu $4,$0,$2
jal 400f98 <malloc>
sw $2,16($30)
\end{verbatim}
\end{varwidth}

\end{figure}

In realloc(), argument registers (\$4 and \$5) are the previous heap address and new size respectively at time of function call. \$2 contains the new heap address after returning from the function. Similarly free(), argument register \$4 contains the base address of the heap block to be freed. Thus, with the content of argument registers and return registers, we can track and differentiate the DMA functions.




\subsection{Variable Record Table (VRT)}




Table \ref{tab:VRT} shows the VRT layout with 3 columns namely, associated bit (1 bit), base address (32 bits), and bound value (8 bits). Each entry of VRT uses 41 bits.  The associated  bit  is  used  to  differentiate  entries  of  a function from other function. As shown in Table \ref{tab:VRT}, top three entries of a function have associated bit different than the function with two entries. With current function entry at top of the table, new function  entry changes  associated  bit to stay different than current function entry. Associative bit is also helpful to link entries to a particular function. During return from the function, all entries with same associative bit on top of the table will be flushed out.


\begin{table}[h]
\begin{center}
\caption{Variable Record Table}
\label{tab:VRT}
\begin{tabular}{|c|c|c|}
\hline
Associated & Variable Address & Bound\\ \hline
\textbf{1} & 0X7FFF60 & 24 \\ \hline
\textbf{1} & 0X7FFF3C & 4 \\ \hline
\textbf{1} & 0X7FFF28 & 4 \\ \hline
0 & 0X7FFE70 & 24 \\ \hline
0 & 0X7FFE60 & 16 \\ \hline
\end{tabular}
\end{center}

\end{table}

\subsubsection{Populating VRT}

During program execution, each jump instruction to a function with local variable or DMA function (malloc()) will populate the VRT. Every new stack address generated using frame pointer will be kept in the table as entry whose bound information is relative to the upper bound of frame pointer unless new entry is populated into the table. Upon next variable entry, we modify the bound information. With DMA function {\tt malloc} we push a new entry into the table. While {\tt realloc} function will search the function entries to match base address and update it with a new base address and bound information. 

\subsubsection{Deleting Entry from VRT}

Table entry need to be flushed out to cope with program requirement. {\tt free()} triggers an entry to be deleted. While return from a function requires deletion of all the entries of the function. Entry with same associated bit from top of the table is deleted in this situation.

\subsection{VRT Overheads}

VRT size is proportional to the number of entries it can accommodate at any time. VRT dynamically grows and shrinks during the program execution. With the deletion of outgoing function's entries VRT relinquish space that is used for new function entries.  

Search overhead for a particular entry can be greatly reduced by concentrating on current function entry. Associated bit helps to differentiate between different functions.

\subsection{Buffer Overflow and VRT} 

Once the local variables' base and bound address are populated in VRT, we can test each array offset and pointer operation to generate the invalid memory address under two representative cases. In the following section, we discuss two different cases of illegal accesses.

\subsubsection{Constant variable index}

Direct access to an array with constant index value beyond the array range can be treated as an out-of-bound access case. If unchecked, such operation will corrupt the out of scope data.  

\begin{figure}[H]
\centering
\begin{varwidth}{\linewidth}
\begin{verbatim}
a[out_of_bound] = 'X'; 
\end{verbatim}
\end{varwidth}
\label{fig:heap_overflow_c_cod}
\end{figure}

Assembly code for array access is shown below. The offset  to the load  ({\tt ld}) instruction produces an address that goes beyond the address space of the variable stored in the entry of VRT. 

\begin{figure}[H]
\centering
\begin{varwidth}{\linewidth}
\begin{verbatim}
4002e0: lw $2,out_of_bound($30) 
\end{verbatim}
\end{varwidth}
\label{fig:heap_oerflow_c_cod}

\end{figure}

\subsubsection{ Loop operation on array or pointer variable} 

This case is common in buffer overflow condition. String library function such as, {\tt strcpy()} overflow during loop. Unchecked increment operation (line 4 of C snippet shown below) on a pointer variable (ptr) produces address beyond the space of variable X.

\begin{figure}[H]
\centering
\begin{varwidth}{\linewidth}
\begin{verbatim}
   1.  char X[6];
   2.  char *ptr = X;
   3.  for(i=0; i<10 ;i++)
   4   ++ptr = '\0';
\end{verbatim}
\end{varwidth}
\label{fig:heap_oeflow_c_cod}
\end{figure}

We can decompose {\tt ++ptr} operation into two instructions in a MIPS like architecture. Firstly, it accesses the variable address and writes it to a register. Secondly, the register is incremented to produce a new address. In the second operation (line 5), \$2 act as source and destination address on which increment operation is done. For an out-of-bound case, \$2 must have addresses that fall in two different entries of VRT. During regular program execution, increment operation must have an address that falls in the same entry of VRT.

\begin{figure}[H]
\centering
\begin{varwidth}{\linewidth}
\begin{verbatim}
 1.    4002e0: lw $2,44($30) 
 2.    4002e8: addu $3,$0,$2
 3.    4002f0: sll $2,$3,0x2  
 4.    4002f8: lw $3,40($30)
 5.    400300: addu $2,$2,$3
\end{verbatim}
\end{varwidth}
\label{fig:heap_oeflow_c_cod}
\end{figure}

 We have used the pipeline micro-architecture in \cite{sah} to implement VRT. Fig. 5 shows the five-stage pipeline architecture with VRT.
 
 \begin{figure}[!h]
\includegraphics[width=\columnwidth]{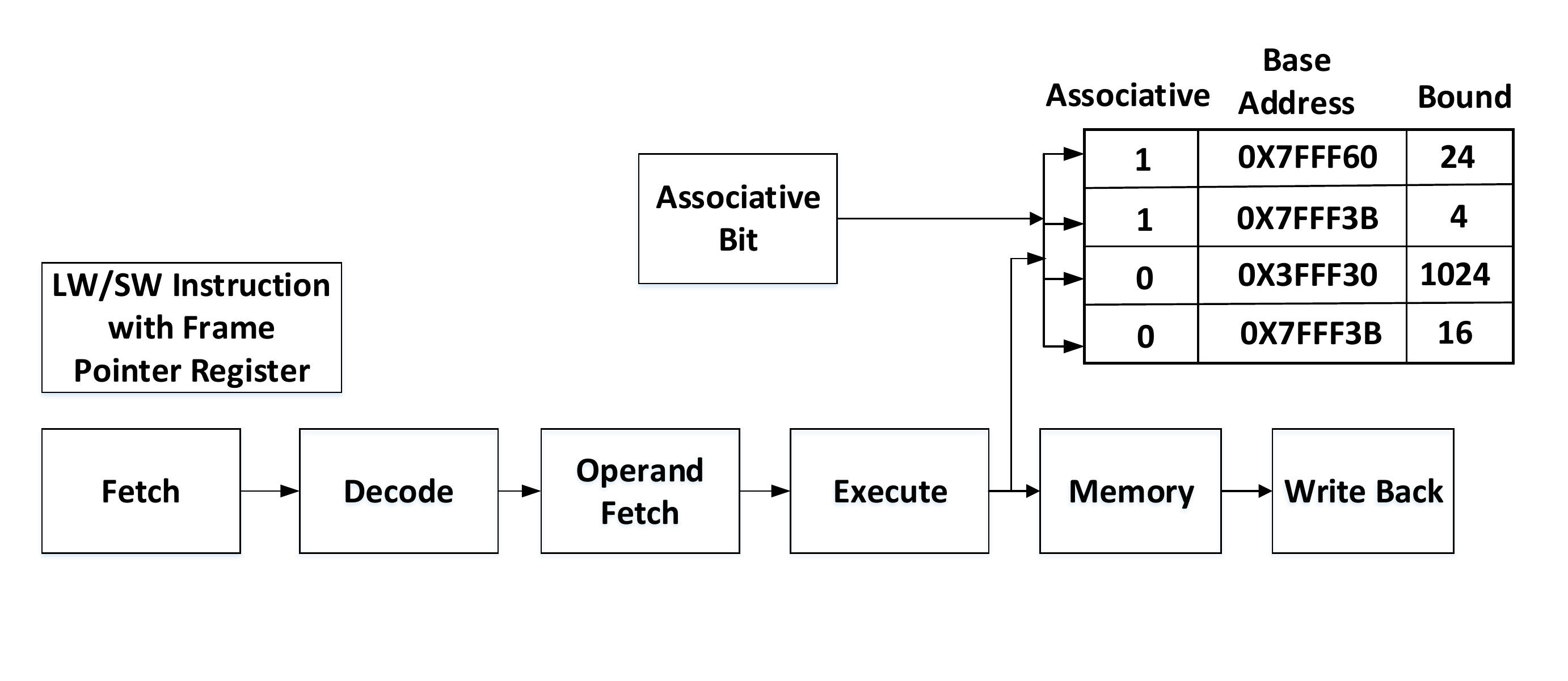}
\caption{Pipeline Architecture to Add Entry }
\label{fig:stack_memory}

\end{figure}

 Process variable table implementation is achieved by augmenting the fetch and execution stage of the processor pipeline with a VRT extraction unit and a memory space unit. Extraction unit checks on the lw and sw instruction with frame pointer registers as an operand. The new address generated in execution stage is stored to the VRT. 
 
 \begin{figure}[!h]
\centering
\includegraphics[width=0.5\textwidth]{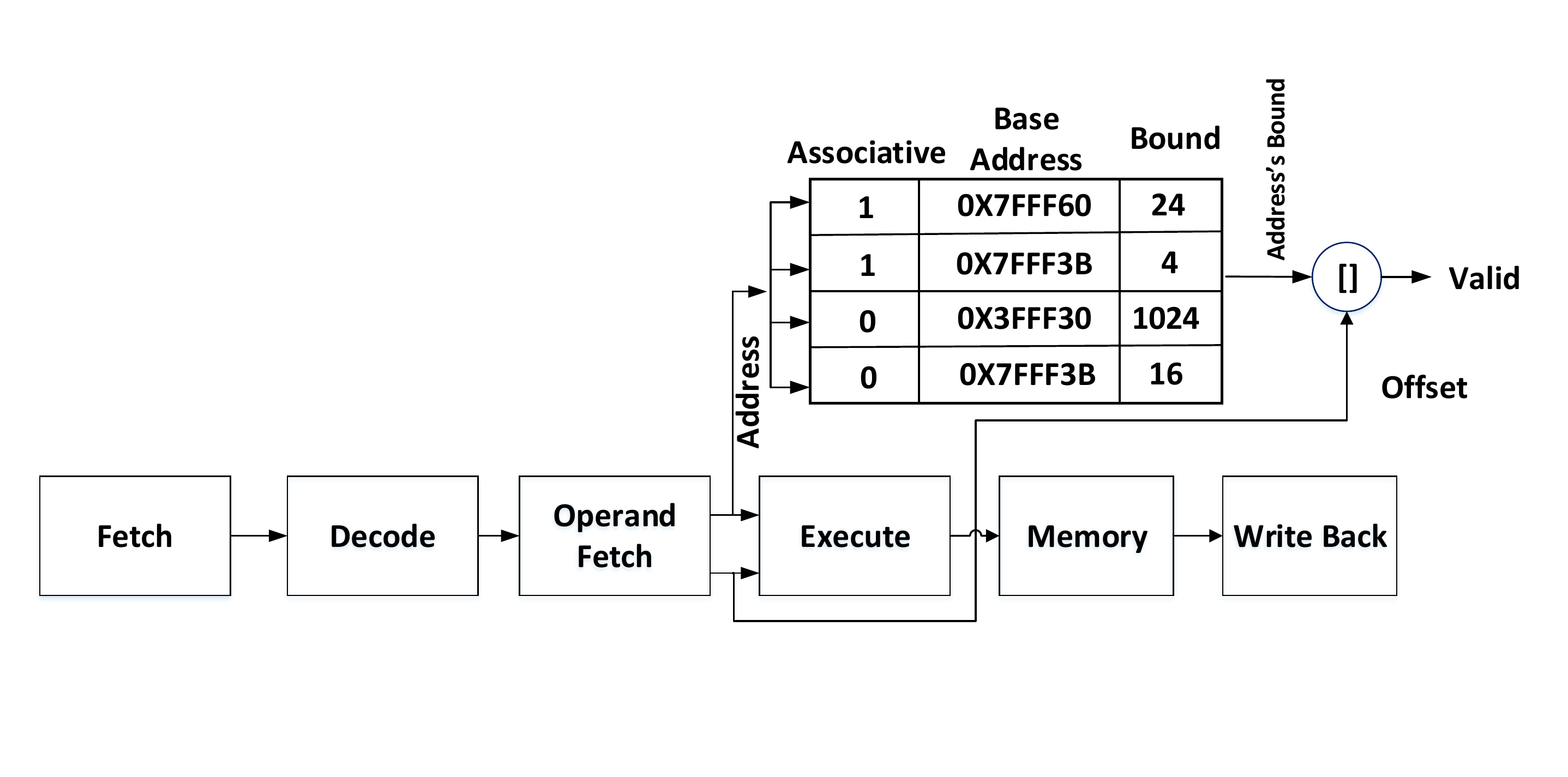}
\vspace{-2ex}
\caption{VRT and Buffer Overflow Check}
\label{fig:shadow}
\end{figure}

Fig. 6 shows overflow detection architecture using VRT. During line 5  operation in above assembly code, VRT gives the bound information of the first operand and is compared with the second operand  If second operand is less than the bound value, we consider it as a within bound access else is an out of bound access.

\section{Experimental Results}
\label{sec:results}

We modified the sim-outorder simulator in SimpleScalar toolset \cite{SimpleScalar} to validate the proposed approach. Sim-outorder simulator is a detailed pipelined micro-architectural simulator in SimpleScalar toolset. It models different runtime parameters in detail with features including instruction profiling, branch prediction, caches, and external memory.  We chose RISC architecture (PISA) architecture with sequential (i.e., in-order) fetch and decode stage maintaining the instruction order. As we rely on an offset value of previous instruction, sequential execution of instruction was important to maintain.

In order to verify our proposed approach, we first populate VRT. We have used MiBench benchmark suite \cite{990739} with  six selected programs for extracting static variable space and six different benchmarks rich in DMA function for heap space operation. Heap and stack space extraction are performed individually. Table \ref{tab:nist-results} shows the static variable and DMA function count respectively for the programs.

\begin{table}[h]
\begin{center}
\caption{Static Variable and DMA Function Count }
\label{tab:nist-results}
\resizebox{\columnwidth}{!}{
\begin{tabular}{|c|c||c|c|c|c|c|}
\hline
Benchmark & \# Variables  & Benchmark & \# malloc()  & \# calloc()  &  \# realloc()  & \# free()  \\ \hline
basicmath & 25 & automotive 	& 10 & 1  & 0 & 5 \\ \hline
bitcount & 49 & consumer    & 49 & 0 & 0 & 41 \\ \hline
qsort & 13 &  network  	& 10 &0  & 0 & 11\\ \hline
CRC32 & 9 & office   & 324 & 71  & 0 & 542\\ \hline
dijkstra & 15 & security & 37 & 9 & 7 & 58 \\ \hline
patricia & 28 & telecomm & 12 & 0 & 0 & 17 \\ \hline
\end{tabular}}
\end{center}
\end{table}

In the case of the office suite of the MiBench, we observe 324 entry to be maximum entries. As one entry of VRT consist of one bit valid bit, 32 bit for base address and 8 bit for bound value altogether use 41 bit per entry resulting in total VRT memory size 324*41 = $\sim$13Kb.

We also implemented VRT on MIT Corpus suite of 290 C programs for buffer overflow case. Each of 290 test cases from MIT Corpus has four different program suffix namely, ok, large, medium, and min. These programs consist of different buffer overflow attributes \cite{Kartkiewicz2009}. For each case, we successfully detect overflow. The instruction count for each class of program is shown in Table \ref{tab:overflow_result}.

\begin{table}[H]
\begin{center}
\caption{Results MIT Corpus C Benchmarks 290 Programs }
\label{tab:overflow_result}
\begin{tabular}{|c|c|c|}
\hline
MIT Corpus Program Class & Instruction Count(Avg.) & Attack Detected?   
\\ \hline
ok   & 18467 & Yes \\ \hline
large  & 18642  & Yes\\ \hline
medium   & 19378 & Yes\\ \hline
min    & 18875 &  Yes\\ \hline
\end{tabular}
\end{center}
\end{table}

\section{Conclusions and Future Work}
\label{sec:conclude}

We have proposed the variable record table (VRT) approach with zero instruction overhead as a countermeasure for buffer overflow attack. We show frame pointer operation and its role in extracting variable space information. With VRT, we can successfully detect the common form of buffer flow attack. In the future, we plan to use VRT for control flow integrity using the variables for verification.  VRT  can also be a useful tool for smart data prefetcher where the data block can be prefetched based on variable size rather than the block size.

\bibliographystyle{unsrt}
\scriptsize{
\bibliography{bib/HT.bib}}

\begin{thebibliography}{10}

\bibitem{NIST}
{NIST National Vulnerability Database}.
\newblock \url{https://nvd.nist.gov/}.

\bibitem{Smashing}
Aleph One.
\newblock {Smashing the stack for fun and profit}.
\newblock \url{Phrack Magazine, 49(14), Nov. 1996.
  http://www.phrack.org/archives/49/P49-14}.

\bibitem{Flawfinder}
{Flawfinder}.
\newblock \url{https://www.dwheeler.com/flawfinder/}.

\bibitem{RATS}
{RATS}.
\newblock
  \url{https://security.web.cern.ch/security/recommendations/en/codetools/rats.shtml/}.

\bibitem{libsafe}
Arash Baratloo, Timothy Tsai, and Navjot Singh.
\newblock {Libsafe: Protecting Critical Elements of Stacks}, 1999.

\bibitem{30x}
W.~M.~J. Richard and H.~J.~K. Paul.
\newblock {Backwards-Compatible Bounds Checking for Arrays and Pointers in {C}
  Programs}.
\newblock In {\em {AADEBUG}}, pages 13--26, 1997.

\bibitem{nonex}
{ Non-Executable User Stack}.
\newblock \url{http://www.false.com/security/linux-stack/.}

\bibitem{ASLR}
G.~S. Kc, A.~D. Keromytis, and V.~Prevelakis.
\newblock {Countering Code-injection Attacks with Instruction-set
  Randomization}.
\newblock CCS '03, pages 272--280, New York, NY, USA, 2003. ACM.

\bibitem{StackGuard}
Crispin Cowan, Calton Pu, Dave Maier, Heather Hintony, Jonathan Walpole, Peat
  Bakke, Steve Beattie, Aaron Grier, Perry Wagle, and Qian Zhang.
\newblock Stackguard: Automatic adaptive detection and prevention of
  buffer-overflow attacks.
\newblock SSYM'98, pages 5--5, Berkeley, CA, USA, 1998. USENIX Association.

\bibitem{Kartkiewicz2009}
K.~Kratkiewicz and R.~Lippmann.
\newblock {Using a Diagnostic Corpus of C Programs to Evaluate Buffer Overflow
  Detection by Static Analysis Tools}.
\newblock 10 2009.

\bibitem{SimpleScalar}
T.~Austin, E.~Larson, and D.~Ernst.
\newblock {SimpleScalar: an infrastructure for computer system modeling}.
\newblock {\em Computer}, 35(2):59--67, Feb 2002.

\bibitem{sah}
L.~K. {Sah}, S.~A. {Islam}, and S.~{Katkoori}.
\newblock {An Efficient Hardware-Oriented Runtime Approach for Stack-based
  Software Buffer Overflow Attacks}.
\newblock In {\em 2018 Asian Hardware Oriented Security and Trust Symposium
  (AsianHOST)}, pages 1--6, Dec 2018.

\bibitem{990739}
M.~R. {Guthaus}, J.~S. {Ringenberg}, D.~{Ernst}, T.~M. {Austin}, T.~{Mudge},
  and R.~B. {Brown}.
\newblock Mibench: A free, commercially representative embedded benchmark
  suite.
\newblock In {\em Proceedings of the Fourth Annual IEEE International Workshop
  on Workload Characterization. WWC-4 (Cat. No.01EX538)}, pages 3--14, Dec
  2001.

\end{thebibliography}

\end{document}